\documentclass[3p,times,twocolumn]{elsarticle}

\usepackage{graphicx}
\usepackage{amsmath,amssymb}
\usepackage{epsfig}
\usepackage{subfigure}

\usepackage{ecrc}

\usepackage{epstopdf}

\volume{00}

\firstpage{1}

\journalname{Nuclear Physics B Proceedings Supplement}

\runauth{S. Cao}

\jid{nppp}

\jnltitlelogo{Nuclear Physics B Proceedings Supplement}

\usepackage{amssymb}

\begin{document}

\begin{frontmatter}



\dochead{}

\title{Transport Theory of Heavy Flavor in Relativistic Nuclear Collisions}


\author{Shanshan Cao}

\address{Nuclear Science Division, Lawrence Berkeley National Laboratory, Berkeley, California 94720, USA}

\begin{abstract}
A short overview is presented for the recent progress in the theory of heavy flavor transport in ultra-relativistic nuclear collisions, including a summary of different transport models, their phenomenological results of heavy meson quenching and flow at RHIC and LHC, a possible solution to the $R_\mathrm{AA}$ vs. $v_2$ puzzle and predictions for heavy flavor observables beyond the current measurements. 
\end{abstract}

\begin{keyword}
relativistic nuclear collisions, heavy flavor, transport theory
\end{keyword}

\end{frontmatter}


\section{Introduction}
\label{sec:introduction}

Heavy quarks serve as valuable probes of the transport properties of the quark-gluon plasma (QGP) matter created in relativistic heavy-ion collisions at the Relativistic Heavy-Ion Collider (RHIC) and the Large Hadron Collider (LHC). Because of their large mass, most heavy quarks are produced at the primordial stage of collisions via hard scatterings and then they travel through and interact with the medium with their flavors conserved, and thus observe the full evolution history of the QGP fireballs. Over the past decade, experimental observations at both RHIC and LHC have revealed a great many interesting data of heavy flavor hadrons and their decay leptons, among which the most surprising ones are their small values of the nuclear modification factor $R_\mathrm{AA}$ and large values of the elliptic flow coefficient $v_2$, which are almost comparable to those of light hadrons \cite{Adare:2010de,Adamczyk:2014uip,Abelev:2014ipa}. This seems contradictory to one's earlier expectation of the mass hierarchy of parton energy loss inside the QGP and is known as the ``heavy flavor puzzle". Therefore, it still remains a great challenge to fully understand the heavy flavor dynamics in heavy-ion collisions. This includes not only parton energy loss inside the QGP, but also heavy flavor initial production, hadronization and hadronic interaction. 

In this talk, a brief overview will be provided for the frequently utilized heavy quark transport models. Then their phenomenological results will be presented and compared with experimental data. After that, recent theoretical developments will be discussed, including predictions for the two-particle correlation functions of heavy flavor and medium modification of heavy mesons in proton-nucleus collisions. 

\section{Transport Models of Heavy Flavor in Heavy-Ion Collisions}
\label{sec:theory}

\subsection{Collisional Energy Loss of Heavy Quarks}
\label{subsec:col}

In the most general form, the heavy quark evolution can be described using the Boltzmann equation:
\begin{equation}
\label{eq:Boltzmann}
\left[\frac{\partial}{\partial t}+\frac{p_i}{E_{\vec{p}}}\frac{\partial}{\partial x_i}+F_i \frac{\partial}{\partial p_i} \right] f_Q (t,\vec{x},\vec{p})= C\left[f_Q\right],
\end{equation}
in which the left hand side is the total time derivative of the heavy quark distribution function and the right hand side represents the collision term. Usually two assumptions are applied: first, one may neglect the drift term (the third term), or the mean free force from the QGP medium exerted on heavy quarks; and second, one can integrate or average over the position space of the distribution function and only concentrate on the evolution of the momentum space without considering the second term. With theses two assumptions, only the partial time derivative (the first term) remains on the left hand side. The collision term can be expressed as a subtraction of the loss term from the gain term:
\begin{equation*}
C\left[f_Q\right]=\int d^3 k \left[w(\vec{p}+\vec{k},\vec{k})f_Q(\vec{p}+\vec{k})-w(\vec{p},\vec{k})f_Q(\vec{p})\right],
\end{equation*}
where $w(p,k)$ represents the transition rate of a heavy quark from momentum $p$ to $p-k$ and can be directly calculated from the microscopic scattering cross sections. 

One may simplify the transport equation with further assumptions. For example, in the quasi-elastic scattering process, we can assume the momentum change of heavy quark during its each scattering with a light parton is small ($|\vec{k}|\ll |\vec{p}|$). Then we have
\begin{equation*}
C\left[f_Q\right]\approx \int d^3 k \left(k_i \frac{\partial}{\partial p_i}+\frac{1}{2}k_i k_j\frac{\partial^2}{\partial p_i \partial p_j}\right)w(\vec{p},\vec{k})f_Q(\vec{p}),
\end{equation*}
and the Boltzmann equation is reduced to the Fokker-Planck equation of the distribution function $f_Q(t,\vec{p})$:
\begin{equation}
\label{eq:FokkerPlank}
\frac{\partial}{\partial t}f_Q=\frac{\partial}{\partial p_i}\left\{A_i(\vec{p}) f_Q+\frac{\partial}{\partial p_j}\left[B_{ij}(\vec{p})f_Q\right]\right\}.
\end{equation}

In addition, we may also assume every heavy quark is scattered multiple times during its evolution inside the medium, then the Fokker-Planck equation can be stochastically realized by the Langevin equation:
\begin{eqnarray}
\label{eq:LangevinX}
&&dx_i=\frac{p_i}{E_{\vec{p}}}dt, \\
\label{eq:LangevinP}
&&dp_i=-\eta_D(\vec{p}) p_i dt+\xi_i dt.
\end{eqnarray}
In Eq. (\ref{eq:LangevinP}), the first term is known as the drag term and the second term is related to the thermal random force. One may refer to Ref. \cite{Rapp:2009my} for calculations of the transport coefficients above -- $A_i$, $B_{ij}$, $\eta_D$ and $\xi_i$. It is worth noticing that these two simplifications from the Boltzmann equation to the Fokker-Planck equation and then to the Langevin equation are only valid for the collisional energy loss, or the $2\rightarrow2$ scattering of heavy quarks inside the QGP, but not for their radiative energy loss because the gluon radiation process usually does not satisfy these two assumptions. 

Various transport models have been constructed to study the heavy quark diffusion inside the dense nuclear matter, such as the parton cascade model based on the Boltzmann equation \cite{Molnar:2006ci,Zhang:2005ni,Uphoff:2011ad,Uphoff:2012gb}, the linearized Boltzmann approach coupled to a hydrodynamic background \cite{Gossiaux:2010yx,Nahrgang:2013saa} and the Langevin-based transport models \cite{Moore:2004tg, He:2011qa,Cao:2011et,Cao:2012jt}. In the Boltzmann models, the most important ingredient is evaluating the collision term. For most current studies, only the leading order (LO) diagrams for heavy quark scatterings with light quarks and gluons are considered. The dominant contribution is from the $t$-channel matrices of the $Qg\rightarrow Qg$ and $Qq\rightarrow Qq$ processes whose infrared singularity is usually regulated by introducing the Debye screening mass into the gluon propagator \cite{Gossiaux:2008jv,Uphoff:2011ad,He:2015pra}. For the Langevin equation, all the interactions are encoded in the transport coefficients. One can use perturbative QCD (pQCD) to calculate these coefficients \cite{Moore:2004tg}, but can also go beyond that. For instance, in Refs. \cite{vanHees:2005wb,vanHees:2007me,He:2011qa}, a non-perturbative resonance scattering method has been proposed to calculate the transport coefficients: one may assume heavy-light quark interaction with certain potential and solve the $T$-matrix using the Lippmann-Schwinger equation from which diffusion coefficients can be extracted. Due to the existence of the resonant states, the energy loss is enhanced compared to the pQCD calculation. One can also use the lattice QCD \cite{Ding:2011hr,Banerjee:2011ra,Kaczmarek:2014jga} to calculate the transport coefficients. However, the current uncertainties of lattice calculations are still large and no reliable inputs for transport models are available. There are other treatments of the collisional energy loss of heavy quark such as the parton-hadron-string dynamics model introduced by Ref. \cite{Song:2015sfa}.

\subsection{Radiative Energy Loss of Heavy Quarks}
\label{subsec:rad}

While collisional energy loss alone is successful in describing heavy flavor observables in the low transverse momentum $p_\mathrm{T}$ region where the phase space for the medium-induced gluon radiation is restricted by the large mass of heavy quarks \cite{Dokshitzer:2001zm,Abir:2012pu}, it has been shown insufficient \cite{Cao:2013ita,Cao:2015hia} at high $p_\mathrm{T}$. 

To incorporate gluon radiation into the Boltzmann transport model, one need to evaluate the pQCD diagrams for the $2\rightarrow3$ processes for the collision term. Although a full evaluation is available \cite{Kunszt:1979iy}, the result is tedious and hard to efficiently implement in numerical calculations. For this reason, the Gunion-Bertsch approximation is adopted by Refs. \cite{Gossiaux:2010yx,Fochler:2013epa} that is derived at high energy limit and reproduce the exact calculation of the matrix elements over a wide rapidity range. The LO pQCD calculation does not include the LPM effect due to the coherent scatterings. To mimic this effect in the numerical simulation, Ref. \cite{Uphoff:2014hza} requires that the heavy quark mean free path is larger than the formation time of radiated gluons times an $X$ factor.

The radiative energy loss has also been implemented in the Langevin framework \cite{Cao:2013ita,Cao:2015hia}:
\begin{equation}
\label{eq:modifiedLangevin}
d\vec{p}/dt=-\eta_D(p)\vec{p}+\vec{\xi}+\vec{f}_g.
\end{equation}
The classical Langevin equation is modified such that apart from the drag force and thermal random force, a third term $\vec{f_g}=-d\vec{p}_g/dt$ is introduced to describe the recoil force exerted on heavy quarks while it radiates gluons. The gluon radiation probability and its energy and momentum distribution can be calculated based on the gluon distribution function taken from the higher-twist energy loss formalism \cite{Guo:2000nz,Majumder:2009ge,Zhang:2003wk}:
\begin{eqnarray*}
\label{eq:gluondistribution}
\frac{dN_\mathrm{g}}{dx dk_\perp^2 dt}=\frac{2\alpha_s  P(x)\hat{q} }{\pi k_\perp^4} {\sin}^2\left(\frac{t-t_i}{2\tau_f}\right)\left(\frac{k_\perp^2}{k_\perp^2+x^2 M^2}\right)^4,
\end{eqnarray*}
in which $x$ is the fractional energy taken by the emitted gluon from its parent heavy quark, and $k_\perp$ is its  transverse momentum. $P(x)$ is the gluon splitting function and $\tau_f={2Ex(1-x)}/{(k_\perp^2+x^2M^2)}$ is the formation time of the gluon with $E$ and $M$ being the energy and mass of heavy quarks. The gluon transport coefficient $\hat{q}$ is related to the heavy quark momentum space diffusion coefficient -- defined in  $\langle\xi^i(t)\xi^j(t')\rangle=\kappa\delta^{ij}\delta(t-t')$ -- via $\hat{q}=2\kappa C_A/C_F$ which is then linked to the drag coefficient through $\eta_D(p)=\kappa/(2TE)$. Therefore only one free parameter remains in Eq. (\ref{eq:modifiedLangevin}). Note that the higher-twist calculation has recently been developed in Refs. \cite{Qin:2014mya,Abir:2015hta} to also incorporate the drag induced radiation.

It has been shown in Ref. \cite{Cao:2015hia} that in 2.76~TeV central Pb-Pb collisions, quasi-elastic scattering dominates the energy loss of heavy quarks with low initial energy while gluon radiation dominates the high energy region. The crossing points are around 7~GeV for charm quark and 18~GeV for bottom quark. This indicates that including both energy loss mechanisms is necessary to study the heavy quark phenomenology at high $p_\mathrm{T}$ as observed at the LHC experiment.

\subsection{Hadronization and Hadronic Interactions}
\label{subsec:hadron}

To fully understand heavy flavor observables, studying parton energy loss alone is not enough. After heavy quarks travel outside the QGP medium, they hadronize into color neutral bound states. For the QGP fireball itself, we can use the standard cooper-frye formalism to sample light hadrons out of it. For heavy quarks, we need to develop a hybrid model of fragmentation plus heavy-light quark coalescence to calculate their hadronization process. High $p_\mathrm{T}$ heavy quarks tend to fragment directly into hadrons. One may use either a proper fragmentation function to calculate the corresponding hadron spectra or use \textsc{Pythia} to simulate this process. On the other hand, it is more probable for lower $p_\mathrm{T}$ heavy quarks to combine with thermal partons from the medium to form new hadrons. This mechanism can be described using either an instantaneous coalescence model \cite{Oh:2009zj,Gossiaux:2009mk,Cao:2015hia} or a resonance recombination model \cite{vanHees:2005wb,vanHees:2007me,He:2011qa}.

It has been shown in Refs. \cite{He:2011qa,Cao:2015hia} that while fragmentation dominates the high $p_\mathrm{T}$ region of heavy quark hadronization, coalescence significantly enhances the heavy quark production rate at medium $p_\mathrm{T}$. For this reason, the coalescence mechanism could generate the bump structure of the $D$ meson $R_\mathrm{AA}$. In addition, coalescence also enhances the $D$ meson $v_2$ since it adds the momentum space anisotropy of light partons onto heavy quarks when $D$ mesons form.

After hadronization, $D$ mesons continue being scattered inside the hadron gas. Two approaches have been used to model these interactions. For example, since now we have both soft hadrons from the QGP fireball and heavy mesons from the heavy quarks, as long as the scattering cross sections between them are known, we can put them into a Boltzmann based cascade model such as the UrQMD \cite{Bass:1998ca} to simulate their further evolution \cite{Cao:2015hia}. An alternative approach is calculating the diffusion coefficient of $D$ mesons inside a hadron gas first and then put it into the Langevin equation \cite{He:2014cla}. It has been shown in Ref. \cite{Cao:2015hia} that due to the additional scatterings of $D$ mesons inside the hadron gas, its $R_\mathrm{AA}$ is further suppressed at high $p_\mathrm{T}$ and its $v_2$ is enhanced by another $20\sim30$\%.

\begin{figure}[tb]
   \epsfig{file=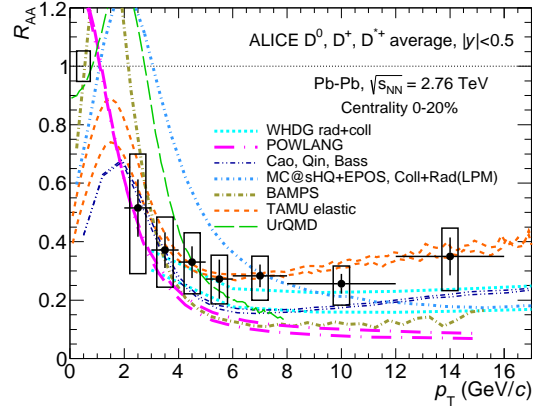, width=0.45\textwidth, clip=}
   \caption{(Color online) The $D$ meson $R_\mathrm{AA}$ in central Pb-Pb collisions at LHC \cite{Abelev:2014ipa}.}
    \label{fig:ALICE_RAA}
\end{figure}

\begin{figure}[tb]
   \epsfig{file=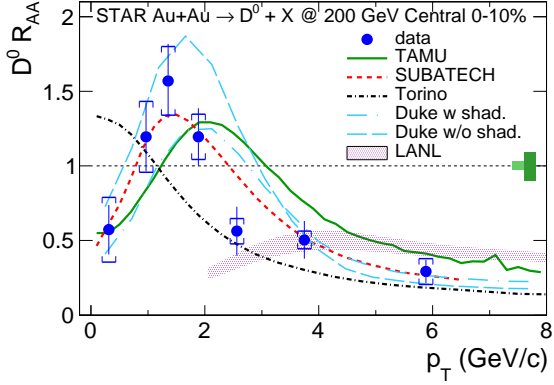, width=0.45\textwidth, clip=}     
   \caption{(Color online) The $D$ meson $R_\mathrm{AA}$ in central Au-Au collisions at RHIC \cite{Adamczyk:2014uip}.}
    \label{fig:RHIC_RAA}
\end{figure}

\begin{figure}[tb]
   \epsfig{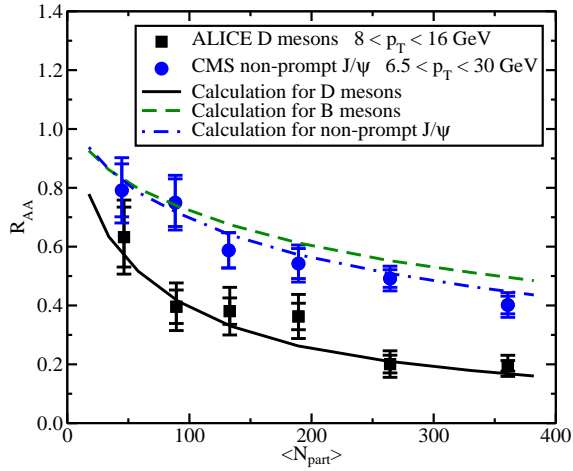}
   \caption{(Color online) The mass dependence of the heavy meson suppression \cite{Cao:2015hia}.}
   \label{fig:massDep}
\end{figure}

\section{Heavy Meson Suppression and Flow}
\label{sec:RAAv2}

Figure \ref{fig:ALICE_RAA} and \ref{fig:RHIC_RAA} summarize the comparison of the $p_\mathrm{T}$ dependence of the $D$ meson suppression between experimental data and model calculations. With proper tunings of the transport coefficients, most models are able to provide reasonable descriptions of the data. As discussed earlier, the bump structure of the $D$ meson $R_\mathrm{AA}$ especially observed at RHIC mainly results from the coalescence process in hadronization. Figure \ref{fig:massDep} shows the participant number dependence of the integrated heavy meson suppression, compared between $D$ meson, $B$ meson and non-prompt $J/\psi$ decayed from $B$-meson. Due to the larger mass of $b$ quark, $B$ meson is less suppressed than $D$ meson. And Fig. \ref{fig:massDep} provides a verification of the mass hierarchy of the parton energy loss inside the QGP from both theoretical and experimental sides.

\begin{figure}[tb]
   \epsfig{file=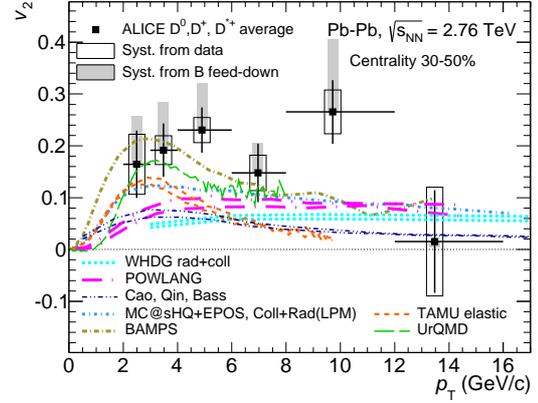, width=0.45\textwidth, clip=}
   \caption{(Color online) The $D$ meson $v_2$ in 30-50\% Pb-Pb collisions at LHC \cite{Abelev:2014ipa}.}
   \label{fig:ALICE_v2}
\end{figure}

Figure \ref{fig:ALICE_v2} compares the $D$ meson $v_2$ between model calculations and measurement at the LHC. A closer investigation together with Fig. \ref{fig:ALICE_RAA} indicates that it still remains a great challenge in theoretical calculations to simultaneously describe $R_\mathrm{AA}$ and $v_2$ exactly. When $R_\mathrm{AA}$ is fixed, $v_2$ is underestimated; and when $v_2$ is fixed, $R_\mathrm{AA}$ is over-suppressed. Recently, there are two studies aiming at this $R_\mathrm{AA}$ vs. $v_2$ puzzle. In Ref. \cite{Das:2015ana}, four different model calculations of the temperature dependence of the drag coefficient are compared and it is suggested that when the $R_\mathrm{AA}$ is fixed, the stronger the drag coefficient is at low temperature, or around $T_\mathrm{c}$, the larger the $v_2$ will be. The physical picture is that if the interaction around $T_\mathrm{c}$ is stronger, heavy quarks will lose most of their energy at later time when the strong anisotropic flow of the QGP has been developed and therefore pick up a larger $v_2$ from the medium. A similar conclusion has been drawn in Ref. \cite{Xu:2014tda} in which a semi-quark-gluon monopole plasma model is introduced that increases the gluon transport coefficient $\hat{q}$ around $T_\mathrm{c}$ and consequently enhances the $v_2$ of hard probe particles.

The study of gluon contribution to heavy meson suppression \cite{Cao:2015kvb} has also been presented at the conference. It has been pointed out that although the gluon splitting process contributes a sizable fraction of the final $D$ meson yield, the influence on the nuclear modification of the single heavy meson production is quite modest due to the limited time for the hard gluons to interact with the dense medium before splitting into heavy quark pairs. On the other hand, the contribution of hard gluons to the nuclear modification of heavy flavor tagged jets has been studied in Refs. \cite{Huang:2013vaa,Huang:2015mva} and shown important.

\section{Observables Beyond the Current Measurements}
\label{sec:newObservables}

Apart from the single particle spectra of heavy flavor, the two-particle correlation functions related to heavy mesons have been actively discussed recently \cite{Zhu:2007ne,Nahrgang:2013saa,Uphoff:2013rka,Cao:2014pka,Cao:2015cba} and shown possible to reveal additional information about heavy flavor dynamics. 

\begin{figure}[tb]
  \epsfig{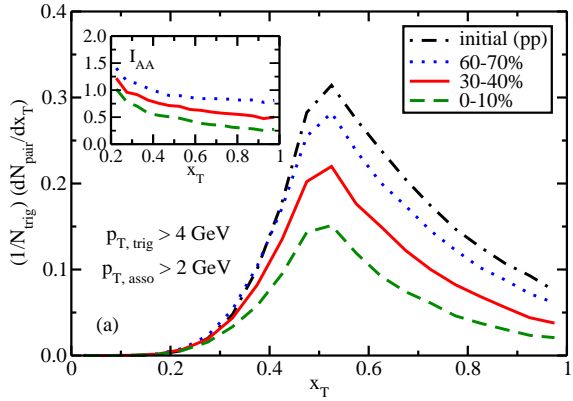}
  \caption{(Color online) The $x_\mathrm{T}$ distribution of $D\overline{D}$ pairs for different centralities \cite{Cao:2015cba}.}
    \label{fig:pAsym}
\end{figure}

In Fig. \ref{fig:pAsym} we study the transverse momentum imbalance of $D\overline{D}$ pairs in heavy-ion collisions. For each event, we select the $D$ or $\overline{D}$ meson that has the highest transverse momentum as the leading (trigger) meson. On the back side, we look for its anti-particle with the highest transverse momentum and select it as the subleading (associated) meson. An angular cut for the away-side subleading meson -- $|\phi_{\rm asso} - \phi_{\rm trig}| \ge 2\pi/3$ -- is applied which should help reduce the background of uncorrelated $D$ and $\overline{D}$'s. The transverse momentum imbalance is defined as $x_\mathrm{T}=p_\mathrm{T,asso}/p_\mathrm{T,trig}$ and its event distribution is shown for different centralities and compared with the proton-proton baseline. Two observations can be found in Fig. \ref{fig:pAsym}: as one moves from proton-proton collisions to more and more central Au-Au collisions, (1) there exist a smaller number of $D\overline{D}$ pairs per triggered event, and (2) the distribution shifts to smaller $x_\mathrm{T}$ (i.e., larger momentum imbalance). These both result from stronger energy loss of heavy quarks inside the QGP. In the subfigure, we also present the ratios between nucleus-nucleus collisions and proton-proton collisions, i.e., $I_\mathrm{AA}$. The values of $I_\mathrm{AA}$ are above unity at small $x_\mathrm{T}$ but below at large $x_\mathrm{T}$. And at large $x_\mathrm{T}$, $I_\mathrm{AA}$ decreases for more central collisions.

\begin{figure}[tb]
   \epsfig{file=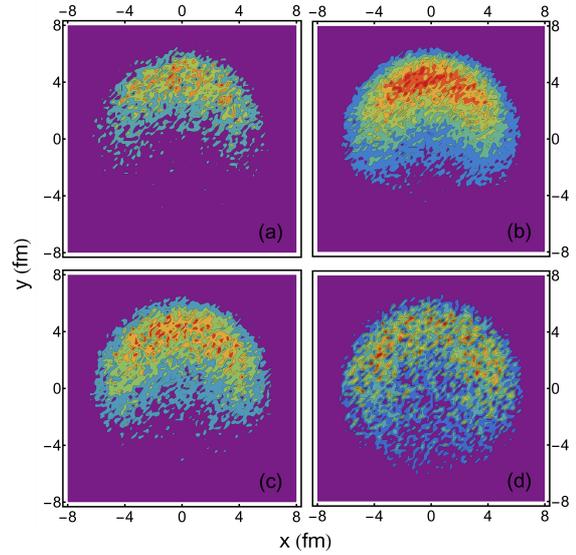, width=0.45\textwidth, clip=}
   \caption{(Color online) The density distribution of the $c\bar{c}$ production positions ($x_\mathrm{init}$, $y_\mathrm{init}$) in 0-10\% Au-Au collisions at RHIC \cite{Cao:2015cba}. The triggered $D$ or $\overline{D}$'s are taken along the out-of-plane directions ($|\phi_\mathrm{trig}-\pi/2| < \pi/6$), and $p_\mathrm{T, trig}>4$~GeV and $p_\mathrm{T, asso}>2$~GeV are implemented. The values of $x_\mathrm{T}$ of the final $D\overline{D}$ pairs are taken as: (a) $x_\mathrm{T}\in[0.2, 0.4]$, (b) $x_\mathrm{T}\in[0.4, 0.6]$, (c) $x_\mathrm{T}\in[0.6, 0.8]$, and (d) $x_\mathrm{T}\in[0.8, 1.0]$.}
    \label{fig:initXY-xT}
\end{figure}

The momentum imbalance not only helps quantify the energy loss of heavy quarks, but also provides us the possibility to probe specific regions of the QGP fireballs. In Fig. \ref{fig:initXY-xT} we investigate the correlation between $x_\mathrm{T}$ of the final $D\overline{D}$ pairs and the initial production positions of the $c\overline{c}$ pairs. We observe that smaller $x_\mathrm{T}$ corresponds to events initially produced at the edge of the QGP fireballs in which one heavy quark travels outside the medium without much interaction while its partner traverse the whole QGP fireball and loses significant amount of energy. On the other hand, larger $x_\mathrm{T}$ corresponds to the initial $c\overline{c}$ pairs that are spread out smoothly over the QGP. 

Apart from the momentum imbalance, one may also study the angular correlation function of heavy meson pairs. It has been pointed out in Refs. \cite{Nahrgang:2013saa,Cao:2015cba} that when yielding similar $D$ meson $R_\mathrm{AA}$, collisional energy loss is more effective in smearing out the away side peak of the $D-\overline{D}$ angular correlation function and enhancing the near side peak due to the radial flow effect of the QGP. Thus if future experiments can measure such correlation functions of heavy flavor pairs, they may provide us with a better understanding of the detailed energy loss mechanisms of heavy quarks.

Another interesting study presented at this conference is the first evaluations of the nuclear modification of heavy quarks in small systems created in proton-nucleus collisions \cite{Nardi:2015pca,Xu:2015iha}. One may first tune a hydrodynamic model so that it provides reasonable descriptions of the soft charged hadron spectra, and then study the evolution of heavy quarks inside this small system with a well-controlled transport model. It has been shown that a significant amount of $D$ meson suppression could be observed in central p-Pb collisions and non-zero $v_2$ has also been predicted for different centrality regions.

\section{Summary}
\label{sec:summary}

In this talk, different transport models and their implementations to heavy quark energy loss in QGP have been summarized. It has been shown that while collisional energy loss dominates at low $p_\mathrm{T}$, radiative energy loss is important at high $p_\mathrm{T}$.  Numerical results of heavy meson $R_\mathrm{AA}$ and $v_2$ have been compared with experimental data and a possible solution to the $R_\mathrm{AA}$ vs. $v_2$ puzzle has been discussed: a closer investigation of the temperature dependence of the interaction strength. Some predictions for future experiments have also been presented, such as the two-particle correlation functions of heavy flavor pairs and the non-trivial nuclear modification of heavy flavor spectra in small systems produced by proton-nucleus collisions.

\section*{Acknowlegements}

This work is funded by the Director, Office of Energy Research, Office of High Energy and Nuclear Physics, Division of Nuclear Physics, of the U.S. Department of Energy under Contract No. DE-AC02-05CH11231, and within the framework of the JET Collaboration.




\bibliographystyle{h-physrev5}
\bibliography{SCrefs}







\end{document}